\documentclass[%
 reprint,
 amsmath,amssymb,
 aps,
]{revtex4-1}

\usepackage{graphicx}
\usepackage{dcolumn}
\usepackage{bm}
\usepackage{color}

\begin{document}

\preprint{AIP/123-QED}

\title{Laser-induced surface relief nanocrowns as a manifestation of nanoscale Rayleigh-Plateau hydrodynamic instability}

\author{D.V. Pavlov$^{1,2}$}
\author{S.O. Gurbatov$^{1,2}$}
\author{S.I. Kudryashov$^{3,4,5}$}
\author{E.L. Gurevich$^{6}$}
\author{A.A. Kuchmizhak$^{1,2}$}

\affiliation{$^1$Far Eastern Federal University, Vladivostok, Russia}
\affiliation{$^2$Institute of Automation and Control Processes, Far Eastern Branch, Russian Academy of Sciences, Vladivostok 690041, Russia}
\affiliation{$^3$Lebedev Physical Institute, Russian Academy of Sciences, 119991 Moscow, Russia}
\affiliation{$^4$National Research Nuclear University MEPhI, 115409 Moscow, Russia}
\affiliation{$^5$ITMO University, 197101 St. Petersburg, Russia}
\affiliation{$^6$Applied Laser Technologies, Ruhr-Universit\"at Bochum, Universit\"atsstra\ss{}e 150, 44801, Bochum, Germany}

\email{gurevich@lat.rub.de}
\email{alex.iacp.dvo@mail.ru}

\date{\today}

\begin{abstract}
Nanoscale hydrodynamic instability of ring-like molten rims around ablative microholes produced in nanometer-thick gold films by tightly focused nanosecond-laser pulses was experimentally explored in terms of laser pulse energy and film thickness. These parametric dependencies of basic instability characteristics - order and period of the resulting nanocrowns - were analyzed, revealing its apparently Rayleigh-Plateau character, as compared to much less consistent possible van der Waals and impact origins. Along with fundamental importance, these findings will put forward pulsed laser ablation as an alternative facile inexpensive table-top approach to study such hydrodynamic instabilities developing at nanosecond temporal and nanometer spatial scales.% The enlightened understanding of such laser-induced instabilities is expected to allow substantial improvement of the direct pulsed laser nanofabrication process regarding more clean and reproducible laser printing of optical filters, transparent conductive coating and biosensing elements.
\end{abstract}

% insert suggested PACS numbers in braces on next line

\maketitle %\maketitle must follow title, authors, abstract and \pacs

% Body of paper goes here. Use proper sectioning commands.
% References should be done using the \cite, \ref, and \label commands

%\section{\label{sec:level1}Introduction}

When a drop or a stone impacts smooth liquid surface, a sequence of transient mm-scale patterns like e.g., jets and crownlets can be observed at microsecond timescale before the energy of impact is dissipated \cite{splashes}. The underlying fluid dynamics behind the formation of such patterns are currently intensively studied in multiple papers and explained in terms of development of a certain type of hydrodynamic instability (Rayleigh-Plateau (R.-P.), Rayleigh-Taylor, Richtmeyer-Meshkov, etc.) \cite{splashes}.

Similar multiphysics phenomena are known to emerge at much faster nanosecond timescales upon either a femto- or nanosecond pulsed laser irradiation of solid films of variable thickness. In particular, ultra-thin films being exposed by laser radiation can become unstable even at temperatures well below melting temperature and undergo dewetting into spherical-like droplets resulting from minimization of the surface energy \cite{trice2007pulsed,wu2010breakup,fowlkes2011self}. Also, laser-induced nanojets reported more the decade ago by Nakata and Chichkov \cite{nakata2003nano,korte2004formation} are widely studied as promising direct laser-pattering technology allowing to produced unique surface morphologies via nanoscale hydrodynamic mass transport. Moreover, the R.-P. instability of liquid laser-induced nanojets is widely explored for either laser-induced forward- (LIFT) or backward transfer (LIBT) of isolated nanoparticles onto acceptor substrate \cite{unger2012time,zywietz2014laser,visser2015toward,zenou2015laser,feinaeugle2018printing}.

Alongside with the significant fundamental interest driving related studies of laser-induced  hydrodynamics at nanoscale \cite{kondic2019liquid}, there are plenty of practical applications of nano-textured films where detailed understanding of complex multiphysics phenomena is motivated by further optimization of the fabrication protocols towards advanced multi-functional devices (see recent reviews \cite{makarov2017light,ye2018dewetting,ruffino2019nanostructuration}). Unfortunately, ultrafast time- and sub-micron length scales characterizing the laser-driven hydrodynamics in metal films can not be captured and visualized directly at appropriate temporal and spatial resolution, allowing only observation of these phenomena via studying resolidified energy-dependent surface morphologies by scanning electron microscopy (SEM). Indeed, pump laser energy being absorbed and converted into heat by conduction-band electrons within a small volume of the metal film dissipates via much faster lateral heat conduction. On the other hand, interaction of the laser-melted material with a substrate as well as other various multiphysics effects further complicates the comprehensive theoretical analysis.

Among others, observation of crownlet formation (i.e. crowning of the molten rim around a through holes) upon direct pulsed laser ablation of metal films was also reported in several papers \cite{nakata2009liquidly,kulchin2014formation,kuchmizhak2015laser}, leaving albeit the question regarding origin of this interesting phenomenon open. In this Letter, we have performed parametric experimental studies of laser-induced ``nanocrown'' formation under single nanosecond-pulse ablation of thin noble-metal films. Our experimental data systematized over a wide range of input experimental parameters (metal film thickness and chemical composition, applied pulse energy, etc.) reveal, for the first time, the crowning process as a direct consequence of the nanoscale R.-P. instability developing in the molten rim. %Alongside with fundamental importance of the these results, our findings substantiate pulsed-laser excitation as an alternative facile inexpensive laboratory method to study the hydrodynamic instabilities developing at nanosecond timescale and nanoscale length scale.

%________________________________Fig21
\begin{figure}
\includegraphics[width=1\columnwidth]{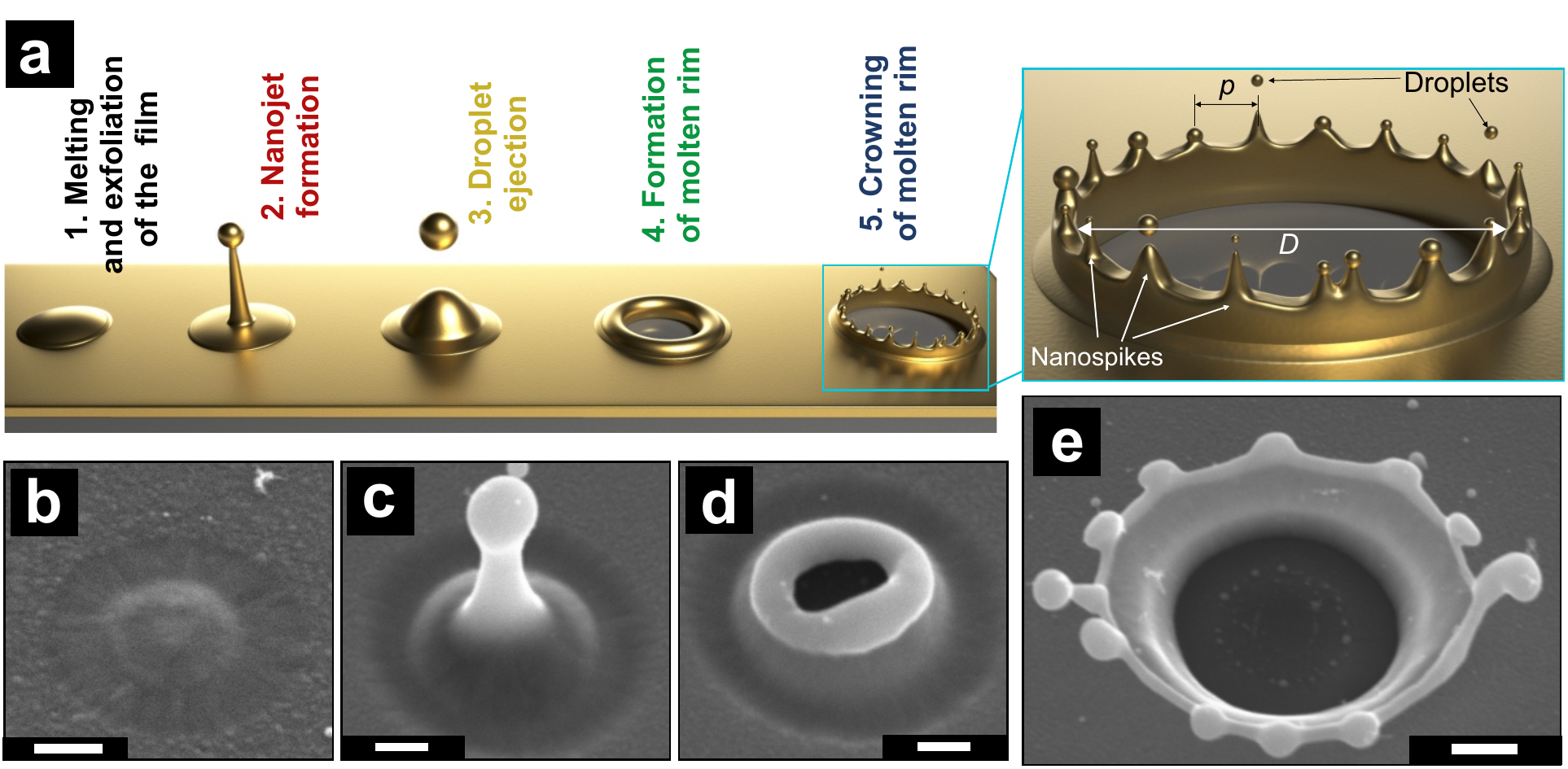}
\caption{\textbf{Crowning process of noble-metal film.} (a) Consequent evolution steps preceding the formation of molten rim and its development to the nanocrown during nanostructuring of a noble metal film on a glass substrate by a single tightly focused ns pulse at increased pulse energy. Inset illustrates a crown with its main characteristic dimensions - diameter $D$, and modulation period $p$. (b-e) Side-view (at an angle of 45$^{\circ}$) SEM images of surface nanostructures produced in glass-supported 120-nm thick Ag film by tightly focused 7-ns laser pulses. The scale bars is 200 nm.}
\end{figure}
%________________________________Fig21

In this study, we used silver film targets widely used for various optical, plasmonic and electronic applications. Silver films of variable thickness ranging from 15 to 120 nm were coated onto a silica glass (or crystalline Si) substrates using e-beam evaporation procedure (Kurt Lesker). The targets were irradiated by linearly-polarized second-harmonic ($\lambda$=532 nm) 7-ns pulses generated by a Nd:YAG laser system at 20-Hz maximal repetition rate (Quantel Brio GRM). Laser beam profile was filtered through a pinhole to provide highly symmetrical laser spot having smooth Gaussian-shaped lateral profile at the focal plane of dry microscope objectives with numerical aperture (NA) of 0.65. The estimated focal spot size on the target surface was R$_{opt}$=0.61$\lambda$(NA)$^{-1}\approx$0.5 $\mu$m. Each surface structure shown in this paper was produced under single-pulse irradiation in air, with its final morphology carefully visualized and analyzed using scanning electron microscopy (Carl Zeiss, Ultra 50+).

Figure 1a schematically illustrates all main relevant steps which ``thermally thin'' noble metal film irradiated with a tightly focused laser pulse proceeds prior to formation of a molten rim and its subsequent evolution into a well-developed nanocrown with periodically arranged nanospikes (``ejecta''). Besides the last step, this sequence of processes and related surface morphologies generally resemble those for fs-pulse irradiation of the same target \cite{korte2004formation,nakata2003nano,unger2012time,kuchmizhak2016laser,naghilou2019femtosecond}. Briefly, under the action of tightly focused laser pulse metal film melts and subsequently exfoliates from the substrate due to stress relaxation caused by the lateral expansion of the metal film, and/or via evaporation process at the film-substrate interface \cite{meshcheryakov2006thermoelastic,demaske2010ablation,inogamov2016solitary,wang2017laser}. The later is more likely for longer nanosecond laser pulses. At increased energy $E$, microscopic hydrodynamic flows pull the molten film toward the optical spot center, yielding in a nanojet. Higher energy leads to development of the R.-P. hydrodynamic instability in the liquid nanojet breaking up its upper end and resulting in vertical ejection of a spherical droplets \cite{rayleigh1878instability}.

At further increase of the pulse energy, formation of sub-micron hole with a pronounced molten rim is observed resulting from a collapse of the bump walls upon build-up of vapor recoil pressure at the ``film-substrate'' interface and/or ejection of the multiple droplets \cite{moening2009formation,kuchmizhak2016ion}. Finally, at some critical diameter $D$, the molten rim accumulating enough material also becomes unstable resulting in periodic modulation of the rim height. Development of a certain mode with the highest growing rate is expected to define an averaged number of nanospikes $N$ as well as a periodic modulation period $p$ for a fixed laser excitation conditions and metal film geometry (see inset of Fig.1a). At the latest step, the nanospikes also supporting the R.-P. instability eject nano-droplets. Each step in the mentioned sequence of events is illustrated by a series of energy-resolved SEM images in Fig.1a showing resolidified surface morphologies produced at increased energy on the 120-nm thick Ag film surface.
Observed nanoscale crowning of the rim around microholes produced in the thin metal films under their single-pulse ablation with a ns-laser pulse generally looks very similar to transient impact of a symmetric micro-object onto a liquid surface (water crowns/splashes) indicating similar nature of the occurring hydrodynamic instability modulating the rim (see for example, \cite{Zhang2010}). However, phase transitions and processes lasting up to sub-microsecond timescales \cite{qi2016time,fang2017direct,jiang2018electrons}, as well as transient behavior of temperature and surface tension gradients significantly complicate the physical picture of such laser-driven crowning.

To get some insight into the features of this process we have systematically studied the effect of laser excitation and film thickness on the rim crowning of the resolidified surface structures. Series of SEM images in Fig.2a shows the result of single-pulse ablation of the 35-nm thick Ag film at increasing pulse energy $E$. These images clearly indicate that no crowning is observed at small diameters of the structure produced at near-threshold energies (fluences) $E\approx E_{th}$. Increase of the applied pulse energy expectedly leads to spreading of the molten rim and appearance of the periodically arranged nanospikes, which number $N$ increases with $E$. Noteworthy, the rim diameter $D$ increases with $E$ demonstrating both typical linear dependence in the $D^2$-$[ln(E)]$ coordinates at fixed metal film thickness $h$ and slight raise of the linear slope (effective energy deposition diameter $D_{th}$) at $h$=120 nm (see Fig. 2b), which is also in agreement with the previously published data \cite{kulchin2014formation,qi2016time}.

The results showing the nanospike number $N$ as a function of nanocrown diameter $D$ obtained for variable Ag film thicknesses $h$ ranging from 15 to 120 nm are summarized in Fig. 2c. These results clearly demonstrate increase of the averaged number of the nanospikes in the rim ($N_{aver}$ shown as linear fits of the colored markers) with $D$ for each tested film thickness as well as increase of $N$ with $h$. The averaged rim modulation period calculated as $p$=$\pi$D/N$_{aver}$ taking into account average number of nanospikes N$_{aver}$ in the rim shows slow increase with the rim diameter $D$ as well as more pronounced linear dependence versus the root of the film thickness $\sqrt{h}$ (see Fig. 3a,b). As can be seen both parameters can be readily used to predictively tune the nanocrown period $p$ over the range between 250 and 750 nm.

 %To the contrary, the focusing conditions (see Inset in Fig.~\ref{Fig22}(a)) and metal film material composition (see Fig.S2 in the Supplementary material) were found to weakly affect the N value.

%________________________________Fig2
\begin{figure}
\includegraphics[width=1\columnwidth]{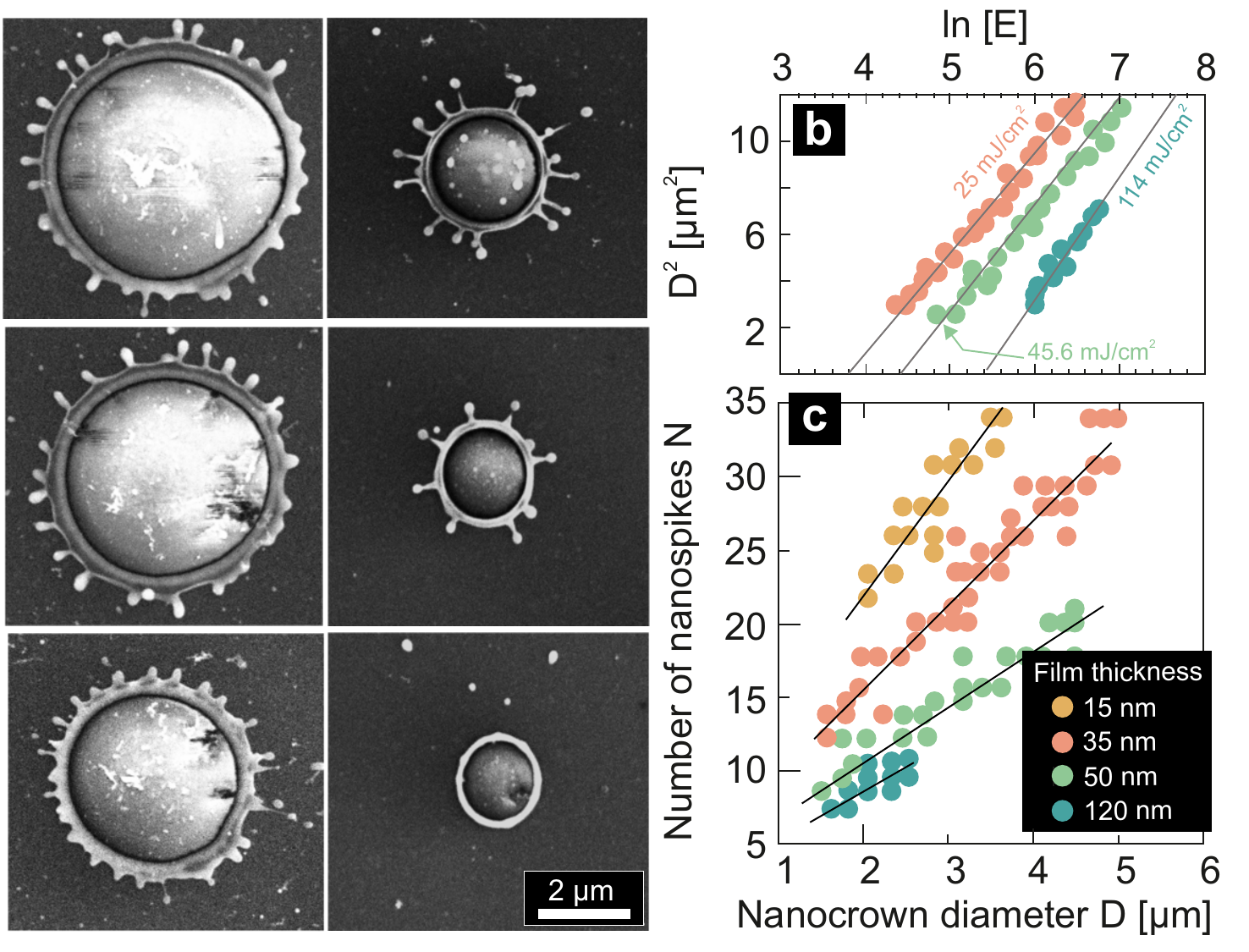}
\caption{\textbf{Crowning of glass-supported Ag films upon their ns-pulse ablation.} (a) Normal-view SEM images illustrating the liquid-phase crowning of the rim formed around the through holes produced by irradiation of 35-nm thick glass-supported Ag film by single 7-ns laser pulse at increased pulse energy $E$. Scale bar in all images corresponds to 2 $\mu$m. (b) Squared nanocrown diameter $D^2$ versus the natural logarithm of the applied pulse energy $ln[E]$ ($E$
in nJ). The linear slope of the fitting line indicates the characteristic energy deposition diameter of $D_{th}\approx$ 2.09$\pm$0.1 $\mu$m. The threshold absorbed fluence $F_{th}$ is shown for each Ag film thickness considering Ag absorbance coefficient A=0.02 \cite{babar2015optical}. (c) Number of nanospikes in the nanocrown $N$ versus the nanocrown diameter $D$ measured at Ag film thicknesses $h$=15, 35, 50 and 120 nm (colored circles).  All data sets are approximated with line fits (solid lines) determining the average number of the nanospikes N$_{aver}$ for a given film thickness $h$.
}
\end{figure}
%________________________________Fig22

%Also noteworthy, the number of nanospikes in the nanocrown was found to be affected by the the substrate type demonstrating more pronounced increase of $N$ versus $D$  resulting in $\approx$ 1.5 fold smaller modulation periods $p$ for Ag film covering thermally conducting Si substrate being compared to those on the bulk glass. To the contrary, focusing conditions (\green{see Fig.S2 of the Supporting Information}) and metal film material composition were found to weakly affect $N$ and $p$ value. \red{These data can be removed in no convincing explanation will be provided...}

Finally, evident crowning of the molten rim under fs-laser pulse irradiation was rarely observed experimentally \cite{nakata2009liquidly,kuchmizhak2015laser}. Anyway, the existing experimental data of fs-laser pulse excitation allows to assess the modulation period $p$ versus the nanocrown diameter $D$ for a 50 nm gold film on silicon with 5 nm intermediate Cr layer shown in Fig. 3a by red stars markers. The data indicates for a certain range of nanocrown diameters $D$ almost twice shorter crowning periods $p$ of the structures produced under fs-pulse irradiation being compared with similar results produced on the similar 50-nm thick Ag film under ns-pulse excitation. Possible reason behind this situation appears to be caused by faster resolidification rate as well as rather high separation velocity of the molten nanobump from the substrate at laser fluences required to produce a through hole initiating crowning \cite{wang2017laser}. Such regime causes the molten material to be ejected in the vertical direction, while the nanobump walls resolidify almost perpendicularly with respect to the substrate, which is observed in multiple experiments \cite{korte2004formation,nakata2003nano,pavlov201910}. %\blue{Moreover, elastic properties, which play an important role for laser-induced microflow formation \cite{meshcheryakov2006thermoelastic}, are different for gold and silver.} %\green{We did not see strong difference between crowning of Au and Ag films under ns-pulse irradiation, possibly not the case for fs pulses. Also to be considered is the difference in adhesion of the film to the substrate with/without Cr adhesion layer. This feature potentially explains the difference in crowning process for glass/Si substrate. } %\revcut{???Additionally, comparison of fs and ns ablation regimes appears to indicate that liquid-phase crowning of the metal films requires relative long heating (developing) times and/or slower velocities of the molten flows supported by relatively long ns laser pulses.}}

Whereas formation mechanisms of the nanobumps and nanojets are intensively discussed in the literature \cite{meshcheryakov2006thermoelastic,ivanov2008mechanism,wang2017laser}, physics of the laser-induced crowning of the rim formed around through hole has never been addressed, to the best of our knowledge. Here, we suggest and compare with our experimental findings three following possible formation scenario: (i) dewetting of the thin films induced by Van der Waals forces; (ii) adopted theory of the liquid drop impact on the liquid surface; (iii) R.-P. theory adopted for the laser melting of thin films. The main points of these models are summarized below.

    (i)  Molten film instability due to {\it Van der Waals interaction} with the substrate \cite{Vrij1966}. For the Van der Waals potential, the grow rate of the instability $\gamma$ can be written as \cite{Oron}:
    \begin{equation}\label{VdW}
        \gamma= \left(\frac{A}{6\pi h}-\frac{\sigma h^3 k^2}{3}\right)\frac{k^2}{\mu},
    \end{equation}
%$$ \gamma= \left(\frac{A}{6\pi h}-\frac{\sigma h^3 k^2}{3}\right)\frac{k^2}{\mu}.$$
where $\mu$ and $\sigma$ are the viscosity and surface tension of the melt respectively. For most of liquid metals the Hamaker constant $A\sim10^{-20}-10^{-21}\,J$ as it is shown in Ref.\cite{Hamaker}.
The critical wave number of the instability $k_c$ limits the range of wave numbers, for which $\gamma >0$. The unstable wave numbers are between $k_{min}=0$ and $k_c=\sqrt{\frac{A}{2\pi\sigma h^4}}\approx 4\times10^5\,1/m$. Hence the wavelength of the observed instability must be larger than $\lambda_{min}\sim 1\,mm$. This estimate is several orders of magnitude larger than the crownlet periods observed in the experiments. Moreover, the period of the dewetting pattern must scale as $p\propto h^2$ (see, for example \cite{seemann2001gaining}), which is not observed in the experiments (see e.g., Fig. 3b). Hence, dewetting due to Van der Waals forces cannot be responsible for the crownlet formation.

(ii) {\it Crownlet formation due to drop impact.} Although the impact of liquid drops on a liquid surface is studied for more than one century \cite{splashes}, the physical mechanisms of the pattern formation are still under discussion \cite{Krechetnikov2009,Zhang2010}. The similarity of the patterns (crownlets and jets) appeared in both systems suggests comparison between the experimental observations of the crownlets generated upon the drop impact and the laser-induced crowning of the metal films. However, as it was mentioned above, there is a significant difference in the experimental conditions: in our experiments the energy for the crown generation is supplied by the incident laser light. Moreover, at the initial step laser-irradiated metal film can be considered as a two-dimensional system interacting with a substrate, whereas the falling drops interact with the volume of the liquid.

Experiments on the drop impact can be analyzed based on the Weber number of the film \cite{Krechetnikov2009}, which is $\mathcal{W}\!e_{\rm film}=\dfrac{\rho v^2h}{\sigma}$, where $\rho$ is the density of the liquid and $v$ - the velocity. Both parameters are difficult to assess experimentally; however, their product is the kinetic energy density of the ejected liquid volume, which must be proportional to the laser fluence $F$ divided by the absorption length $\ell_{\rm abs}$, which is about 20\,nm for noble metals. So, we can write that $\rho v^2 \propto F/\ell_{\rm abs}$ and hence $\mathcal{W}\!e_{\rm film}\propto F$. Krechetnikov and Homsy \cite{Krechetnikov2009} observed that the number of crown spikes linearly \textit{decreases} with the Weber number; however, our experiments suggest that for all probed film thicknesses $h$ the averaged number of nanospikes $N_{aver}$ \textit{increases} with the fluence which is proportional to nanocrown diameter $D$, in its turn (see Fig. 2b). Hence, we conclude that it is not possible to use similarity between these systems to explain the crownlets in our laser ablation experiments.

%________________________________Fig3
\begin{figure}
\includegraphics[width=1.0\columnwidth]{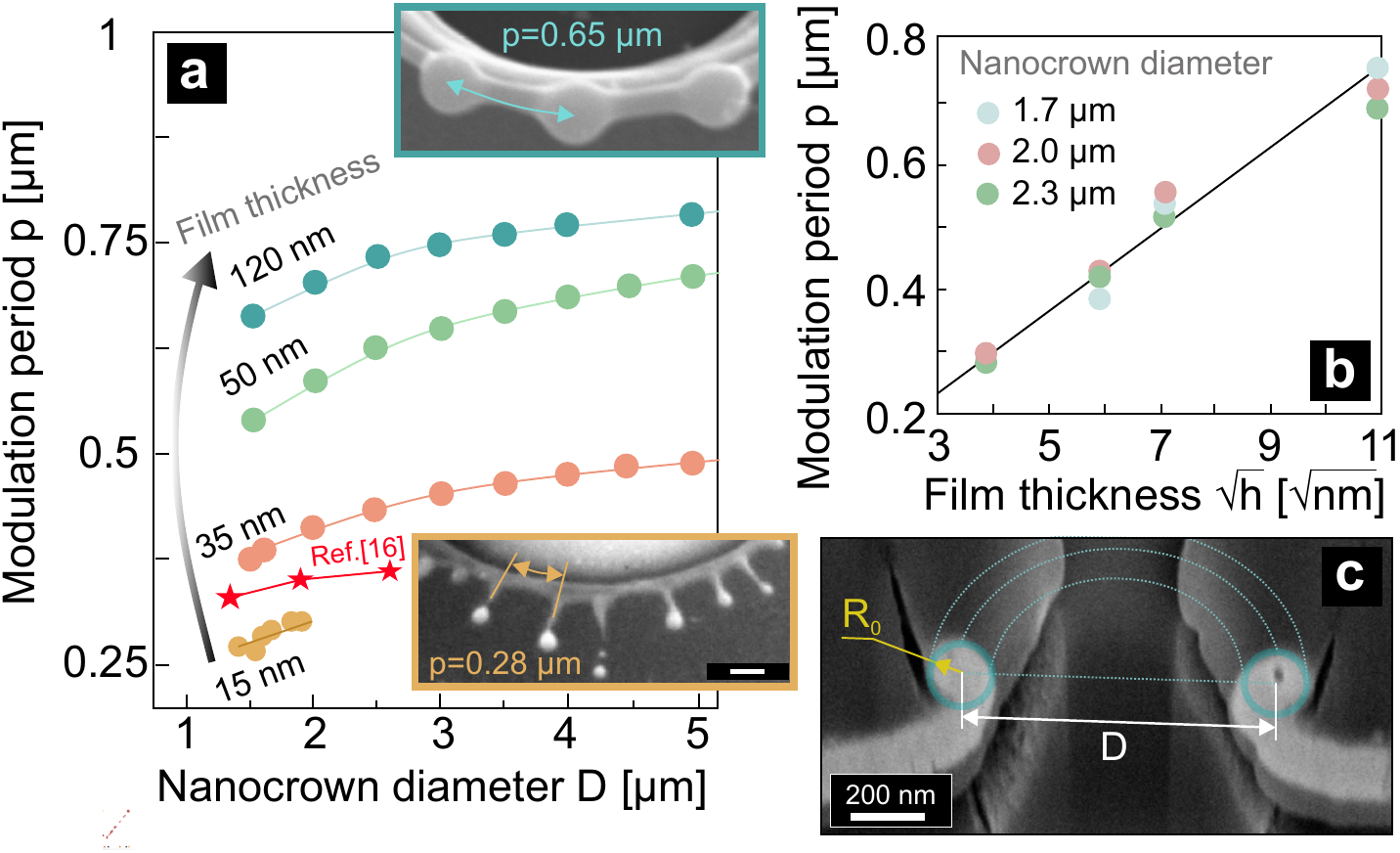}% Here is how to import EPS art
\caption{\textbf{Characteristics of laser-induced crowning.} (a) Modulation period $p$ versus the nanocrown diameter $D$ measured for various film thicknesses $h$. %ranging from 15 to 120 nm.
%Fitting curves are provided as eye guides.
The red stars provide the data %$p$($D$) dependence calculated using the experimental data
from Ref.\cite{nakata2009liquidly}. Inset SEM images illustrate difference in the nanocrown period at $h$=120 and 15 nm; Scale bars are 200 nm.
(b) Modulation period $p$ as a function of the root of the metal film thickness $\sqrt{h}$ given for several representative nanocrown diameters $D$. %= 1.7, 2 and 2.3 $\mu$m. %The linear fit is provided as an eye guide.
(c) Side-view (view angle of 60$^o$) SEM image of central FIB cross-section cut of through hole produced in 120-nm thick glass-supported Ag film under single pulse irradiation. %The characteristic rim size $R_0$ and nanocrown diameter $D$ are indicated.
}
\end{figure}
%________________________________Fig3

(iii) {\it The Rayleigh-Plateau instability.} Crownlets formed upon drop impact in water were studied in the context of the R.-P. instability under strong assumptions that the liquid rim is not connected to the substrate and has a cylindrical shape \cite{Zhang2010}. In order to describe crownlets in thin laser-ablated films and achieve more accurate predictions, we use the theory of the R.-P. instability developed in our previous publications for a liquid filament in a channel \cite{khare2007dewetting,seemann2011wetting}. Taking the wedge angle of the channel $\Psi=0$, which corresponds to the plane surface, we obtain the following equation for the period of the fastest growing mode of the instability:
\begin{equation}\label{RPfgm}
    p=\dfrac{\sqrt{2}\pi R_0}{\sin\theta}\sqrt{1-\cos\theta},
\end{equation}
%$$\Lambda=\dfrac{\sqrt{2}\pi w}{\sin\theta}\sqrt{1-\cos\theta},$$
where $\theta$ is the contact angle of liquid metal on the substrate and $R_0$ is the size of the molten rim.

To assess the rim size $R_0$, using focused ion beam (FIB) milling we prepared a cross-section cut of the through hole produced in a 120-nm thick Ag film at pulse energy slightly above the corresponding ablation threshold to avoid strong modulation of the rim thickness owing to crowning (see Fig. 1d). The Ag film was protected with a 300-nm thick Ti overlayer prior to FIB milling process. Side-view SEM image of the produced central cut is shown in Fig. 3c revealing the characteristic size of the rim around $R_0$=80$\pm$15 nm. Taking into account the contact angle of noble metal films on a quartz surface $\theta\approx 140^o$ for further estimations \cite{ricci2001wetting}, one can achieve the crowning period $p$=720$\pm$150 nm for $h$=120 nm, which agrees well with our experimental findings (see Fig. 3a).
Volume of the molten rim $\pi^2DR_0^2$ should be proportional to the volume of the molten section of the laser-irradiated film $\approx \dfrac{\pi h}{4}D^2$ Diameter of the molten section $D$ can be expressed as $D\approx D_{th}\left(\ln\left[\dfrac{E}{E_{th}}\right]\right)^{0.5}$, where $D_{th}$ and $E_{th}$ are characteristic energy deposition diameter and threshold pulse energy. Both are constant at the fixed film thickness $h$ and can be estimated as a slope of the $D^2$-$[ln(E)]$ dependence and its intersection with X-axis, respectively (see Fig. 2b). This assumption gives the rim size
\begin{equation}
R_{0}\approx \alpha \dfrac{1}{2\sqrt{\pi}}\sqrt{D_{th}h} \cdot \left(\ln\left[\dfrac{E}{E_{th}}\right]\right)^{0.25},
\end{equation}
where $\alpha<$1 accounts for the volume of the molten material elected by spherical droplets. The obtained dependence agrees well with the experimentally observed trends: (i) pronounced linear dependence of the crowning period $p\propto R_0$ versus the square root of the film thickness $\sqrt{h}$ and (ii) more weaker effect of $p$ on the nanocrown diameter $D$ (see Fig. 3a,b).

To conclude, in this Letter via systematic studies of ns-pulse laser ablation of thin noble-metal films of variable thickness we showed that the R.-P. instability developing in the molten rim around the ablative through hole gives a plausible explanation of the observed crowning: predicted modulation period is proportional to the square root of the layer thickness $\sqrt{h}$ and the nanocrown diameter $D$, which fully agrees with the experimental data (Fig.3). Along with fundamental importance of the these results revealing the laser-induced crowning of thin noble-metal foils as a direct consequence of the R.-P. instability, our findings also have evident practical importance. In particular, the results substantiate pulsed laser ablation as an alternative facile inexpensive table-top method to study the hydrodynamic instabilities developing at nanosecond timescale and nanoscale length scale. Moreover, the understanding of the laser-induced instabilities is expected to allow optimization of the pulsed laser nanofabrication process, namely, the more clean and reproducible laser printing of optical filters, transparent conductive coating and biosensors \cite{kuchmizhak2016ion,danilov2016background,paeng2015low,busleev2019fano}. Finally, the controllable crowning accompanied by ejection of the droplets can be considered as a tool for generation of quasi-monodispersed nanoparticles of calibrated size.

This work was supported by Russian Science Foundation (grant no. 19-12-10165). S.I.K. acknowledges the Government of the Russian Federation  through the ITMO Visiting Professorship Program (Grant 074-U01).


\begin{thebibliography}{45}%
\makeatletter
\providecommand \@ifxundefined [1]{%
 \@ifx{#1\undefined}
}%
\providecommand \@ifnum [1]{%
 \ifnum #1\expandafter \@firstoftwo
 \else \expandafter \@secondoftwo
 \fi
}%
\providecommand \@ifx [1]{%
 \ifx #1\expandafter \@firstoftwo
 \else \expandafter \@secondoftwo
 \fi
}%
\providecommand \natexlab [1]{#1}%
\providecommand \enquote  [1]{``#1''}%
\providecommand \bibnamefont  [1]{#1}%
\providecommand \bibfnamefont [1]{#1}%
\providecommand \citenamefont [1]{#1}%
\providecommand \href@noop [0]{\@secondoftwo}%
\providecommand \href [0]{\begingroup \@sanitize@url \@href}%
\providecommand \@href[1]{\@@startlink{#1}\@@href}%
\providecommand \@@href[1]{\endgroup#1\@@endlink}%
\providecommand \@sanitize@url [0]{\catcode `\\12\catcode `\$12\catcode
  `\&12\catcode `\#12\catcode `\^12\catcode `\_12\catcode `\%12\relax}%
\providecommand \@@startlink[1]{}%
\providecommand \@@endlink[0]{}%
\providecommand \url  [0]{\begingroup\@sanitize@url \@url }%
\providecommand \@url [1]{\endgroup\@href {#1}{\urlprefix }}%
\providecommand \urlprefix  [0]{URL }%
\providecommand \Eprint [0]{\href }%
\providecommand \doibase [0]{http://dx.doi.org/}%
\providecommand \selectlanguage [0]{\@gobble}%
\providecommand \bibinfo  [0]{\@secondoftwo}%
\providecommand \bibfield  [0]{\@secondoftwo}%
\providecommand \translation [1]{[#1]}%
\providecommand \BibitemOpen [0]{}%
\providecommand \bibitemStop [0]{}%
\providecommand \bibitemNoStop [0]{.\EOS\space}%
\providecommand \EOS [0]{\spacefactor3000\relax}%
\providecommand \BibitemShut  [1]{\csname bibitem#1\endcsname}%
\let\auto@bib@innerbib\@empty
%</preamble>
\bibitem [{\citenamefont {Worthington}(1908)}]{splashes}%
  \BibitemOpen
  \bibfield  {author} {\bibinfo {author} {\bibfnamefont {A.~M.}\ \bibnamefont
  {Worthington}},\ }\href@noop {} {\emph {\bibinfo {title} {A study of
  splashes}}}\ (\bibinfo  {publisher} {Longmans, Green, and Co.},\ \bibinfo
  {address} {London},\ \bibinfo {year} {1908})\BibitemShut {NoStop}%
\bibitem [{\citenamefont {Trice}\ \emph {et~al.}(2007)\citenamefont {Trice},
  \citenamefont {Thomas}, \citenamefont {Favazza}, \citenamefont
  {Sureshkumar},\ and\ \citenamefont {Kalyanaraman}}]{trice2007pulsed}%
  \BibitemOpen
  \bibfield  {author} {\bibinfo {author} {\bibfnamefont {J.}~\bibnamefont
  {Trice}}, \bibinfo {author} {\bibfnamefont {D.}~\bibnamefont {Thomas}},
  \bibinfo {author} {\bibfnamefont {C.}~\bibnamefont {Favazza}}, \bibinfo
  {author} {\bibfnamefont {R.}~\bibnamefont {Sureshkumar}}, \ and\ \bibinfo
  {author} {\bibfnamefont {R.}~\bibnamefont {Kalyanaraman}},\ }\href@noop {}
  {\bibfield  {journal} {\bibinfo  {journal} {Physical Review B}\ }\textbf
  {\bibinfo {volume} {75}},\ \bibinfo {pages} {235439} (\bibinfo {year}
  {2007})}\BibitemShut {NoStop}%
\bibitem [{\citenamefont {Wu}\ \emph {et~al.}(2010)\citenamefont {Wu},
  \citenamefont {Fowlkes}, \citenamefont {Rack}, \citenamefont {Diez},\ and\
  \citenamefont {Kondic}}]{wu2010breakup}%
  \BibitemOpen
  \bibfield  {author} {\bibinfo {author} {\bibfnamefont {Y.}~\bibnamefont
  {Wu}}, \bibinfo {author} {\bibfnamefont {J.~D.}\ \bibnamefont {Fowlkes}},
  \bibinfo {author} {\bibfnamefont {P.~D.}\ \bibnamefont {Rack}}, \bibinfo
  {author} {\bibfnamefont {J.~A.}\ \bibnamefont {Diez}}, \ and\ \bibinfo
  {author} {\bibfnamefont {L.}~\bibnamefont {Kondic}},\ }\href@noop {}
  {\bibfield  {journal} {\bibinfo  {journal} {Langmuir}\ }\textbf {\bibinfo
  {volume} {26}},\ \bibinfo {pages} {11972} (\bibinfo {year}
  {2010})}\BibitemShut {NoStop}%
\bibitem [{\citenamefont {Fowlkes}\ \emph {et~al.}(2011)\citenamefont
  {Fowlkes}, \citenamefont {Kondic}, \citenamefont {Diez}, \citenamefont {Wu},\
  and\ \citenamefont {Rack}}]{fowlkes2011self}%
  \BibitemOpen
  \bibfield  {author} {\bibinfo {author} {\bibfnamefont {J.~D.}\ \bibnamefont
  {Fowlkes}}, \bibinfo {author} {\bibfnamefont {L.}~\bibnamefont {Kondic}},
  \bibinfo {author} {\bibfnamefont {J.}~\bibnamefont {Diez}}, \bibinfo {author}
  {\bibfnamefont {Y.}~\bibnamefont {Wu}}, \ and\ \bibinfo {author}
  {\bibfnamefont {P.~D.}\ \bibnamefont {Rack}},\ }\href@noop {} {\bibfield
  {journal} {\bibinfo  {journal} {Nano letters}\ }\textbf {\bibinfo {volume}
  {11}},\ \bibinfo {pages} {2478} (\bibinfo {year} {2011})}\BibitemShut
  {NoStop}%
\bibitem [{\citenamefont {Nakata}\ \emph {et~al.}(2003)\citenamefont {Nakata},
  \citenamefont {Okada},\ and\ \citenamefont {Maeda}}]{nakata2003nano}%
  \BibitemOpen
  \bibfield  {author} {\bibinfo {author} {\bibfnamefont {Y.}~\bibnamefont
  {Nakata}}, \bibinfo {author} {\bibfnamefont {T.}~\bibnamefont {Okada}}, \
  and\ \bibinfo {author} {\bibfnamefont {M.}~\bibnamefont {Maeda}},\
  }\href@noop {} {\bibfield  {journal} {\bibinfo  {journal} {Japanese Journal
  of Applied Physics}\ }\textbf {\bibinfo {volume} {42}},\ \bibinfo {pages}
  {L1452} (\bibinfo {year} {2003})}\BibitemShut {NoStop}%
\bibitem [{\citenamefont {Korte}\ \emph {et~al.}(2004)\citenamefont {Korte},
  \citenamefont {Koch},\ and\ \citenamefont {Chichkov}}]{korte2004formation}%
  \BibitemOpen
  \bibfield  {author} {\bibinfo {author} {\bibfnamefont {F.}~\bibnamefont
  {Korte}}, \bibinfo {author} {\bibfnamefont {J.}~\bibnamefont {Koch}}, \ and\
  \bibinfo {author} {\bibfnamefont {B.}~\bibnamefont {Chichkov}},\ }\href@noop
  {} {\bibfield  {journal} {\bibinfo  {journal} {Applied Physics A}\ }\textbf
  {\bibinfo {volume} {79}},\ \bibinfo {pages} {879} (\bibinfo {year}
  {2004})}\BibitemShut {NoStop}%
\bibitem [{\citenamefont {Unger}\ \emph {et~al.}(2012)\citenamefont {Unger},
  \citenamefont {Koch}, \citenamefont {Overmeyer},\ and\ \citenamefont
  {Chichkov}}]{unger2012time}%
  \BibitemOpen
  \bibfield  {author} {\bibinfo {author} {\bibfnamefont {C.}~\bibnamefont
  {Unger}}, \bibinfo {author} {\bibfnamefont {J.}~\bibnamefont {Koch}},
  \bibinfo {author} {\bibfnamefont {L.}~\bibnamefont {Overmeyer}}, \ and\
  \bibinfo {author} {\bibfnamefont {B.~N.}\ \bibnamefont {Chichkov}},\
  }\href@noop {} {\bibfield  {journal} {\bibinfo  {journal} {Optics express}\
  }\textbf {\bibinfo {volume} {20}},\ \bibinfo {pages} {24864} (\bibinfo {year}
  {2012})}\BibitemShut {NoStop}%
\bibitem [{\citenamefont {Zywietz}\ \emph {et~al.}(2014)\citenamefont
  {Zywietz}, \citenamefont {Evlyukhin}, \citenamefont {Reinhardt},\ and\
  \citenamefont {Chichkov}}]{zywietz2014laser}%
  \BibitemOpen
  \bibfield  {author} {\bibinfo {author} {\bibfnamefont {U.}~\bibnamefont
  {Zywietz}}, \bibinfo {author} {\bibfnamefont {A.~B.}\ \bibnamefont
  {Evlyukhin}}, \bibinfo {author} {\bibfnamefont {C.}~\bibnamefont
  {Reinhardt}}, \ and\ \bibinfo {author} {\bibfnamefont {B.~N.}\ \bibnamefont
  {Chichkov}},\ }\href@noop {} {\bibfield  {journal} {\bibinfo  {journal}
  {Nature communications}\ }\textbf {\bibinfo {volume} {5}},\ \bibinfo {pages}
  {3402} (\bibinfo {year} {2014})}\BibitemShut {NoStop}%
\bibitem [{\citenamefont {Visser}\ \emph {et~al.}(2015)\citenamefont {Visser},
  \citenamefont {Pohl}, \citenamefont {Sun}, \citenamefont {R{\"o}mer},
  \citenamefont {Huis in~'t Veld},\ and\ \citenamefont
  {Lohse}}]{visser2015toward}%
  \BibitemOpen
  \bibfield  {author} {\bibinfo {author} {\bibfnamefont {C.~W.}\ \bibnamefont
  {Visser}}, \bibinfo {author} {\bibfnamefont {R.}~\bibnamefont {Pohl}},
  \bibinfo {author} {\bibfnamefont {C.}~\bibnamefont {Sun}}, \bibinfo {author}
  {\bibfnamefont {G.-W.}\ \bibnamefont {R{\"o}mer}}, \bibinfo {author}
  {\bibfnamefont {B.}~\bibnamefont {Huis in~'t Veld}}, \ and\ \bibinfo {author}
  {\bibfnamefont {D.}~\bibnamefont {Lohse}},\ }\href@noop {} {\bibfield
  {journal} {\bibinfo  {journal} {Advanced materials}\ }\textbf {\bibinfo
  {volume} {27}},\ \bibinfo {pages} {4087} (\bibinfo {year}
  {2015})}\BibitemShut {NoStop}%
\bibitem [{\citenamefont {Zenou}\ \emph {et~al.}(2015)\citenamefont {Zenou},
  \citenamefont {Sa'Ar},\ and\ \citenamefont {Kotler}}]{zenou2015laser}%
  \BibitemOpen
  \bibfield  {author} {\bibinfo {author} {\bibfnamefont {M.}~\bibnamefont
  {Zenou}}, \bibinfo {author} {\bibfnamefont {A.}~\bibnamefont {Sa'Ar}}, \ and\
  \bibinfo {author} {\bibfnamefont {Z.}~\bibnamefont {Kotler}},\ }\href@noop {}
  {\bibfield  {journal} {\bibinfo  {journal} {Scientific reports}\ }\textbf
  {\bibinfo {volume} {5}},\ \bibinfo {pages} {17265} (\bibinfo {year}
  {2015})}\BibitemShut {NoStop}%
\bibitem [{\citenamefont {Feinaeugle}\ \emph {et~al.}(2018)\citenamefont
  {Feinaeugle}, \citenamefont {Pohl}, \citenamefont {Bor}, \citenamefont
  {Vaneker},\ and\ \citenamefont {R{\"o}mer}}]{feinaeugle2018printing}%
  \BibitemOpen
  \bibfield  {author} {\bibinfo {author} {\bibfnamefont {M.}~\bibnamefont
  {Feinaeugle}}, \bibinfo {author} {\bibfnamefont {R.}~\bibnamefont {Pohl}},
  \bibinfo {author} {\bibfnamefont {T.}~\bibnamefont {Bor}}, \bibinfo {author}
  {\bibfnamefont {T.}~\bibnamefont {Vaneker}}, \ and\ \bibinfo {author}
  {\bibfnamefont {G.-w.}\ \bibnamefont {R{\"o}mer}},\ }\href@noop {} {\bibfield
   {journal} {\bibinfo  {journal} {AddItive manufacturing}\ }\textbf {\bibinfo
  {volume} {24}},\ \bibinfo {pages} {391} (\bibinfo {year} {2018})}\BibitemShut
  {NoStop}%
\bibitem [{\citenamefont {Kondic}\ \emph {et~al.}(2019)\citenamefont {Kondic},
  \citenamefont {Gonz{\'a}lez}, \citenamefont {Diez}, \citenamefont {Fowlkes},\
  and\ \citenamefont {Rack}}]{kondic2019liquid}%
  \BibitemOpen
  \bibfield  {author} {\bibinfo {author} {\bibfnamefont {L.}~\bibnamefont
  {Kondic}}, \bibinfo {author} {\bibfnamefont {A.~G.}\ \bibnamefont
  {Gonz{\'a}lez}}, \bibinfo {author} {\bibfnamefont {J.~A.}\ \bibnamefont
  {Diez}}, \bibinfo {author} {\bibfnamefont {J.~D.}\ \bibnamefont {Fowlkes}}, \
  and\ \bibinfo {author} {\bibfnamefont {P.}~\bibnamefont {Rack}},\ }\href@noop
  {} {\bibfield  {journal} {\bibinfo  {journal} {Annual Review of Fluid
  Mechanics}\ }\textbf {\bibinfo {volume} {52}} (\bibinfo {year}
  {2019})}\BibitemShut {NoStop}%
\bibitem [{\citenamefont {Makarov}\ \emph {et~al.}(2017)\citenamefont
  {Makarov}, \citenamefont {Zalogina}, \citenamefont {Tajik}, \citenamefont
  {Zuev}, \citenamefont {Rybin}, \citenamefont {Kuchmizhak}, \citenamefont
  {Juodkazis},\ and\ \citenamefont {Kivshar}}]{makarov2017light}%
  \BibitemOpen
  \bibfield  {author} {\bibinfo {author} {\bibfnamefont {S.~V.}\ \bibnamefont
  {Makarov}}, \bibinfo {author} {\bibfnamefont {A.~S.}\ \bibnamefont
  {Zalogina}}, \bibinfo {author} {\bibfnamefont {M.}~\bibnamefont {Tajik}},
  \bibinfo {author} {\bibfnamefont {D.~A.}\ \bibnamefont {Zuev}}, \bibinfo
  {author} {\bibfnamefont {M.~V.}\ \bibnamefont {Rybin}}, \bibinfo {author}
  {\bibfnamefont {A.~A.}\ \bibnamefont {Kuchmizhak}}, \bibinfo {author}
  {\bibfnamefont {S.}~\bibnamefont {Juodkazis}}, \ and\ \bibinfo {author}
  {\bibfnamefont {Y.}~\bibnamefont {Kivshar}},\ }\href@noop {} {\bibfield
  {journal} {\bibinfo  {journal} {Laser \& Photonics Reviews}\ }\textbf
  {\bibinfo {volume} {11}},\ \bibinfo {pages} {1700108} (\bibinfo {year}
  {2017})}\BibitemShut {NoStop}%
\bibitem [{\citenamefont {Ye}\ \emph {et~al.}(2018)\citenamefont {Ye},
  \citenamefont {Zuev},\ and\ \citenamefont {Makarov}}]{ye2018dewetting}%
  \BibitemOpen
  \bibfield  {author} {\bibinfo {author} {\bibfnamefont {J.}~\bibnamefont
  {Ye}}, \bibinfo {author} {\bibfnamefont {D.}~\bibnamefont {Zuev}}, \ and\
  \bibinfo {author} {\bibfnamefont {S.}~\bibnamefont {Makarov}},\ }\href@noop
  {} {\bibfield  {journal} {\bibinfo  {journal} {International Materials
  Reviews}\ ,\ \bibinfo {pages} {1}} (\bibinfo {year} {2018})}\BibitemShut
  {NoStop}%
\bibitem [{\citenamefont {Ruffino}\ and\ \citenamefont
  {Grimaldi}(2019)}]{ruffino2019nanostructuration}%
  \BibitemOpen
  \bibfield  {author} {\bibinfo {author} {\bibfnamefont {F.}~\bibnamefont
  {Ruffino}}\ and\ \bibinfo {author} {\bibfnamefont {M.~G.}\ \bibnamefont
  {Grimaldi}},\ }\href@noop {} {\bibfield  {journal} {\bibinfo  {journal}
  {Nanomaterials}\ }\textbf {\bibinfo {volume} {9}},\ \bibinfo {pages} {1133}
  (\bibinfo {year} {2019})}\BibitemShut {NoStop}%
\bibitem [{\citenamefont {Nakata}\ \emph {et~al.}(2009)\citenamefont {Nakata},
  \citenamefont {Tsuchida}, \citenamefont {Miyanaga},\ and\ \citenamefont
  {Furusho}}]{nakata2009liquidly}%
  \BibitemOpen
  \bibfield  {author} {\bibinfo {author} {\bibfnamefont {Y.}~\bibnamefont
  {Nakata}}, \bibinfo {author} {\bibfnamefont {K.}~\bibnamefont {Tsuchida}},
  \bibinfo {author} {\bibfnamefont {N.}~\bibnamefont {Miyanaga}}, \ and\
  \bibinfo {author} {\bibfnamefont {H.}~\bibnamefont {Furusho}},\ }\href@noop
  {} {\bibfield  {journal} {\bibinfo  {journal} {Applied Surface Science}\
  }\textbf {\bibinfo {volume} {255}},\ \bibinfo {pages} {9761} (\bibinfo {year}
  {2009})}\BibitemShut {NoStop}%
\bibitem [{\citenamefont {Kulchin}\ \emph {et~al.}(2014)\citenamefont
  {Kulchin}, \citenamefont {Vitrik}, \citenamefont {Kuchmizhak}, \citenamefont
  {Emel'Yanov}, \citenamefont {Ionin}, \citenamefont {Kudryashov},\ and\
  \citenamefont {Makarov}}]{kulchin2014formation}%
  \BibitemOpen
  \bibfield  {author} {\bibinfo {author} {\bibfnamefont {Y.~N.}\ \bibnamefont
  {Kulchin}}, \bibinfo {author} {\bibfnamefont {O.}~\bibnamefont {Vitrik}},
  \bibinfo {author} {\bibfnamefont {A.}~\bibnamefont {Kuchmizhak}}, \bibinfo
  {author} {\bibfnamefont {V.}~\bibnamefont {Emel'Yanov}}, \bibinfo {author}
  {\bibfnamefont {A.}~\bibnamefont {Ionin}}, \bibinfo {author} {\bibfnamefont
  {S.}~\bibnamefont {Kudryashov}}, \ and\ \bibinfo {author} {\bibfnamefont
  {S.}~\bibnamefont {Makarov}},\ }\href@noop {} {\bibfield  {journal} {\bibinfo
   {journal} {Physical Review E}\ }\textbf {\bibinfo {volume} {90}},\ \bibinfo
  {pages} {023017} (\bibinfo {year} {2014})}\BibitemShut {NoStop}%
\bibitem [{\citenamefont {Kuchmizhak}\ \emph {et~al.}(2015)\citenamefont
  {Kuchmizhak}, \citenamefont {Pavlov}, \citenamefont {Vitrik},\ and\
  \citenamefont {Kulchin}}]{kuchmizhak2015laser}%
  \BibitemOpen
  \bibfield  {author} {\bibinfo {author} {\bibfnamefont {A.}~\bibnamefont
  {Kuchmizhak}}, \bibinfo {author} {\bibfnamefont {D.}~\bibnamefont {Pavlov}},
  \bibinfo {author} {\bibfnamefont {O.}~\bibnamefont {Vitrik}}, \ and\ \bibinfo
  {author} {\bibfnamefont {Y.~N.}\ \bibnamefont {Kulchin}},\ }\href@noop {}
  {\bibfield  {journal} {\bibinfo  {journal} {Applied Surface Science}\
  }\textbf {\bibinfo {volume} {357}},\ \bibinfo {pages} {2378} (\bibinfo {year}
  {2015})}\BibitemShut {NoStop}%
\bibitem [{\citenamefont {Kuchmizhak}\ \emph
  {et~al.}(2016{\natexlab{a}})\citenamefont {Kuchmizhak}, \citenamefont
  {Vitrik}, \citenamefont {Kulchin}, \citenamefont {Storozhenko}, \citenamefont
  {Mayor}, \citenamefont {Mirochnik}, \citenamefont {Makarov}, \citenamefont
  {Milichko}, \citenamefont {Kudryashov}, \citenamefont {Zhakhovsky} \emph
  {et~al.}}]{kuchmizhak2016laser}%
  \BibitemOpen
  \bibfield  {author} {\bibinfo {author} {\bibfnamefont {A.}~\bibnamefont
  {Kuchmizhak}}, \bibinfo {author} {\bibfnamefont {O.}~\bibnamefont {Vitrik}},
  \bibinfo {author} {\bibfnamefont {Y.}~\bibnamefont {Kulchin}}, \bibinfo
  {author} {\bibfnamefont {D.}~\bibnamefont {Storozhenko}}, \bibinfo {author}
  {\bibfnamefont {A.}~\bibnamefont {Mayor}}, \bibinfo {author} {\bibfnamefont
  {A.}~\bibnamefont {Mirochnik}}, \bibinfo {author} {\bibfnamefont
  {S.}~\bibnamefont {Makarov}}, \bibinfo {author} {\bibfnamefont
  {V.}~\bibnamefont {Milichko}}, \bibinfo {author} {\bibfnamefont
  {S.}~\bibnamefont {Kudryashov}}, \bibinfo {author} {\bibfnamefont
  {V.}~\bibnamefont {Zhakhovsky}},  \emph {et~al.},\ }\href@noop {} {\bibfield
  {journal} {\bibinfo  {journal} {Nanoscale}\ }\textbf {\bibinfo {volume}
  {8}},\ \bibinfo {pages} {12352} (\bibinfo {year}
  {2016}{\natexlab{a}})}\BibitemShut {NoStop}%
\bibitem [{\citenamefont {Naghilou}\ \emph {et~al.}(2019)\citenamefont
  {Naghilou}, \citenamefont {He}, \citenamefont {Schubert}, \citenamefont
  {Zhigilei},\ and\ \citenamefont {Kautek}}]{naghilou2019femtosecond}%
  \BibitemOpen
  \bibfield  {author} {\bibinfo {author} {\bibfnamefont {A.}~\bibnamefont
  {Naghilou}}, \bibinfo {author} {\bibfnamefont {M.}~\bibnamefont {He}},
  \bibinfo {author} {\bibfnamefont {J.~S.}\ \bibnamefont {Schubert}}, \bibinfo
  {author} {\bibfnamefont {L.}~\bibnamefont {Zhigilei}}, \ and\ \bibinfo
  {author} {\bibfnamefont {W.}~\bibnamefont {Kautek}},\ }\href@noop {}
  {\bibfield  {journal} {\bibinfo  {journal} {Physical Chemistry Chemical
  Physics}\ } (\bibinfo {year} {2019})}\BibitemShut {NoStop}%
\bibitem [{\citenamefont {Meshcheryakov}\ and\ \citenamefont
  {Bulgakova}(2006)}]{meshcheryakov2006thermoelastic}%
  \BibitemOpen
  \bibfield  {author} {\bibinfo {author} {\bibfnamefont {Y.~P.}\ \bibnamefont
  {Meshcheryakov}}\ and\ \bibinfo {author} {\bibfnamefont {N.}~\bibnamefont
  {Bulgakova}},\ }\href@noop {} {\bibfield  {journal} {\bibinfo  {journal}
  {Applied Physics A}\ }\textbf {\bibinfo {volume} {82}},\ \bibinfo {pages}
  {363} (\bibinfo {year} {2006})}\BibitemShut {NoStop}%
\bibitem [{\citenamefont {Demaske}\ \emph {et~al.}(2010)\citenamefont
  {Demaske}, \citenamefont {Zhakhovsky}, \citenamefont {Inogamov},\ and\
  \citenamefont {Oleynik}}]{demaske2010ablation}%
  \BibitemOpen
  \bibfield  {author} {\bibinfo {author} {\bibfnamefont {B.~J.}\ \bibnamefont
  {Demaske}}, \bibinfo {author} {\bibfnamefont {V.~V.}\ \bibnamefont
  {Zhakhovsky}}, \bibinfo {author} {\bibfnamefont {N.~A.}\ \bibnamefont
  {Inogamov}}, \ and\ \bibinfo {author} {\bibfnamefont {I.~I.}\ \bibnamefont
  {Oleynik}},\ }\href@noop {} {\bibfield  {journal} {\bibinfo  {journal}
  {Physical Review B}\ }\textbf {\bibinfo {volume} {82}},\ \bibinfo {pages}
  {064113} (\bibinfo {year} {2010})}\BibitemShut {NoStop}%
\bibitem [{\citenamefont {Inogamov}\ \emph {et~al.}(2016)\citenamefont
  {Inogamov}, \citenamefont {Zhakhovsky}, \citenamefont {Khokhlov},
  \citenamefont {Petrov},\ and\ \citenamefont {Migdal}}]{inogamov2016solitary}%
  \BibitemOpen
  \bibfield  {author} {\bibinfo {author} {\bibfnamefont {N.~A.}\ \bibnamefont
  {Inogamov}}, \bibinfo {author} {\bibfnamefont {V.~V.}\ \bibnamefont
  {Zhakhovsky}}, \bibinfo {author} {\bibfnamefont {V.~A.}\ \bibnamefont
  {Khokhlov}}, \bibinfo {author} {\bibfnamefont {Y.~V.}\ \bibnamefont
  {Petrov}}, \ and\ \bibinfo {author} {\bibfnamefont {K.~P.}\ \bibnamefont
  {Migdal}},\ }\href@noop {} {\bibfield  {journal} {\bibinfo  {journal}
  {Nanoscale research letters}\ }\textbf {\bibinfo {volume} {11}},\ \bibinfo
  {pages} {177} (\bibinfo {year} {2016})}\BibitemShut {NoStop}%
\bibitem [{\citenamefont {Wang}\ \emph {et~al.}(2017)\citenamefont {Wang},
  \citenamefont {Kuchmizhak}, \citenamefont {Li}, \citenamefont {Juodkazis},
  \citenamefont {Vitrik}, \citenamefont {Kulchin}, \citenamefont {Zhakhovsky},
  \citenamefont {Danilov}, \citenamefont {Ionin}, \citenamefont {Kudryashov}
  \emph {et~al.}}]{wang2017laser}%
  \BibitemOpen
  \bibfield  {author} {\bibinfo {author} {\bibfnamefont {X.}~\bibnamefont
  {Wang}}, \bibinfo {author} {\bibfnamefont {A.}~\bibnamefont {Kuchmizhak}},
  \bibinfo {author} {\bibfnamefont {X.}~\bibnamefont {Li}}, \bibinfo {author}
  {\bibfnamefont {S.}~\bibnamefont {Juodkazis}}, \bibinfo {author}
  {\bibfnamefont {O.}~\bibnamefont {Vitrik}}, \bibinfo {author} {\bibfnamefont
  {Y.~N.}\ \bibnamefont {Kulchin}}, \bibinfo {author} {\bibfnamefont
  {V.}~\bibnamefont {Zhakhovsky}}, \bibinfo {author} {\bibfnamefont
  {P.}~\bibnamefont {Danilov}}, \bibinfo {author} {\bibfnamefont
  {A.}~\bibnamefont {Ionin}}, \bibinfo {author} {\bibfnamefont
  {S.}~\bibnamefont {Kudryashov}},  \emph {et~al.},\ }\href@noop {} {\bibfield
  {journal} {\bibinfo  {journal} {Physical Review Applied}\ }\textbf {\bibinfo
  {volume} {8}},\ \bibinfo {pages} {044016} (\bibinfo {year}
  {2017})}\BibitemShut {NoStop}%
\bibitem [{\citenamefont {Rayleigh}(1878)}]{rayleigh1878instability}%
  \BibitemOpen
  \bibfield  {author} {\bibinfo {author} {\bibfnamefont {L.}~\bibnamefont
  {Rayleigh}},\ }\href@noop {} {\bibfield  {journal} {\bibinfo  {journal}
  {Proceedings of the London mathematical society}\ }\textbf {\bibinfo {volume}
  {1}},\ \bibinfo {pages} {4} (\bibinfo {year} {1878})}\BibitemShut {NoStop}%
\bibitem [{\citenamefont {Moening}\ \emph {et~al.}(2009)\citenamefont
  {Moening}, \citenamefont {Thanawala},\ and\ \citenamefont
  {Georgiev}}]{moening2009formation}%
  \BibitemOpen
  \bibfield  {author} {\bibinfo {author} {\bibfnamefont {J.~P.}\ \bibnamefont
  {Moening}}, \bibinfo {author} {\bibfnamefont {S.~S.}\ \bibnamefont
  {Thanawala}}, \ and\ \bibinfo {author} {\bibfnamefont {D.~G.}\ \bibnamefont
  {Georgiev}},\ }\href@noop {} {\bibfield  {journal} {\bibinfo  {journal}
  {Applied Physics A}\ }\textbf {\bibinfo {volume} {95}},\ \bibinfo {pages}
  {635} (\bibinfo {year} {2009})}\BibitemShut {NoStop}%
\bibitem [{\citenamefont {Kuchmizhak}\ \emph
  {et~al.}(2016{\natexlab{b}})\citenamefont {Kuchmizhak}, \citenamefont
  {Gurbatov}, \citenamefont {Vitrik}, \citenamefont {Kulchin}, \citenamefont
  {Milichko}, \citenamefont {Makarov},\ and\ \citenamefont
  {Kudryashov}}]{kuchmizhak2016ion}%
  \BibitemOpen
  \bibfield  {author} {\bibinfo {author} {\bibfnamefont {A.}~\bibnamefont
  {Kuchmizhak}}, \bibinfo {author} {\bibfnamefont {S.}~\bibnamefont
  {Gurbatov}}, \bibinfo {author} {\bibfnamefont {O.}~\bibnamefont {Vitrik}},
  \bibinfo {author} {\bibfnamefont {Y.}~\bibnamefont {Kulchin}}, \bibinfo
  {author} {\bibfnamefont {V.}~\bibnamefont {Milichko}}, \bibinfo {author}
  {\bibfnamefont {S.}~\bibnamefont {Makarov}}, \ and\ \bibinfo {author}
  {\bibfnamefont {S.}~\bibnamefont {Kudryashov}},\ }\href@noop {} {\bibfield
  {journal} {\bibinfo  {journal} {Scientific reports}\ }\textbf {\bibinfo
  {volume} {6}},\ \bibinfo {pages} {19410} (\bibinfo {year}
  {2016}{\natexlab{b}})}\BibitemShut {NoStop}%
\bibitem [{\citenamefont {Zhang}\ \emph {et~al.}(2010)\citenamefont {Zhang},
  \citenamefont {Brunet}, \citenamefont {Eggers},\ and\ \citenamefont
  {Deegan}}]{Zhang2010}%
  \BibitemOpen
  \bibfield  {author} {\bibinfo {author} {\bibfnamefont {L.~V.}\ \bibnamefont
  {Zhang}}, \bibinfo {author} {\bibfnamefont {P.}~\bibnamefont {Brunet}},
  \bibinfo {author} {\bibfnamefont {J.}~\bibnamefont {Eggers}}, \ and\ \bibinfo
  {author} {\bibfnamefont {R.~D.}\ \bibnamefont {Deegan}},\ }\href {\doibase
  10.1063/1.3526743} {\bibfield  {journal} {\bibinfo  {journal} {Physics of
  Fluids}\ }\textbf {\bibinfo {volume} {22}},\ \bibinfo {pages} {122105}
  (\bibinfo {year} {2010})},\ \Eprint
  {http://arxiv.org/abs/https://doi.org/10.1063/1.3526743}
  {https://doi.org/10.1063/1.3526743} \BibitemShut {NoStop}%
\bibitem [{\citenamefont {Qi}\ \emph {et~al.}(2016)\citenamefont {Qi},
  \citenamefont {Paeng}, \citenamefont {Yeo}, \citenamefont {Kim},
  \citenamefont {Wang}, \citenamefont {Chen},\ and\ \citenamefont
  {Grigoropoulos}}]{qi2016time}%
  \BibitemOpen
  \bibfield  {author} {\bibinfo {author} {\bibfnamefont {D.}~\bibnamefont
  {Qi}}, \bibinfo {author} {\bibfnamefont {D.}~\bibnamefont {Paeng}}, \bibinfo
  {author} {\bibfnamefont {J.}~\bibnamefont {Yeo}}, \bibinfo {author}
  {\bibfnamefont {E.}~\bibnamefont {Kim}}, \bibinfo {author} {\bibfnamefont
  {L.}~\bibnamefont {Wang}}, \bibinfo {author} {\bibfnamefont {S.}~\bibnamefont
  {Chen}}, \ and\ \bibinfo {author} {\bibfnamefont {C.~P.}\ \bibnamefont
  {Grigoropoulos}},\ }\href@noop {} {\bibfield  {journal} {\bibinfo  {journal}
  {Applied Physics Letters}\ }\textbf {\bibinfo {volume} {108}},\ \bibinfo
  {pages} {211602} (\bibinfo {year} {2016})}\BibitemShut {NoStop}%
\bibitem [{\citenamefont {Fang}\ \emph {et~al.}(2017)\citenamefont {Fang},
  \citenamefont {Vorobyev},\ and\ \citenamefont {Guo}}]{fang2017direct}%
  \BibitemOpen
  \bibfield  {author} {\bibinfo {author} {\bibfnamefont {R.}~\bibnamefont
  {Fang}}, \bibinfo {author} {\bibfnamefont {A.}~\bibnamefont {Vorobyev}}, \
  and\ \bibinfo {author} {\bibfnamefont {C.}~\bibnamefont {Guo}},\ }\href@noop
  {} {\bibfield  {journal} {\bibinfo  {journal} {Light: Science \&
  Applications}\ }\textbf {\bibinfo {volume} {6}},\ \bibinfo {pages} {e16256}
  (\bibinfo {year} {2017})}\BibitemShut {NoStop}%
\bibitem [{\citenamefont {Jiang}\ \emph {et~al.}(2018)\citenamefont {Jiang},
  \citenamefont {Wang}, \citenamefont {Li}, \citenamefont {Cui},\ and\
  \citenamefont {Lu}}]{jiang2018electrons}%
  \BibitemOpen
  \bibfield  {author} {\bibinfo {author} {\bibfnamefont {L.}~\bibnamefont
  {Jiang}}, \bibinfo {author} {\bibfnamefont {A.-D.}\ \bibnamefont {Wang}},
  \bibinfo {author} {\bibfnamefont {B.}~\bibnamefont {Li}}, \bibinfo {author}
  {\bibfnamefont {T.-H.}\ \bibnamefont {Cui}}, \ and\ \bibinfo {author}
  {\bibfnamefont {Y.-F.}\ \bibnamefont {Lu}},\ }\href@noop {} {\bibfield
  {journal} {\bibinfo  {journal} {Light: Science \& Applications}\ }\textbf
  {\bibinfo {volume} {7}},\ \bibinfo {pages} {17134} (\bibinfo {year}
  {2018})}\BibitemShut {NoStop}%
\bibitem [{\citenamefont {Babar}\ and\ \citenamefont
  {Weaver}(2015)}]{babar2015optical}%
  \BibitemOpen
  \bibfield  {author} {\bibinfo {author} {\bibfnamefont {S.}~\bibnamefont
  {Babar}}\ and\ \bibinfo {author} {\bibfnamefont {J.}~\bibnamefont {Weaver}},\
  }\href@noop {} {\bibfield  {journal} {\bibinfo  {journal} {Applied Optics}\
  }\textbf {\bibinfo {volume} {54}},\ \bibinfo {pages} {477} (\bibinfo {year}
  {2015})}\BibitemShut {NoStop}%
\bibitem [{\citenamefont {Pavlov}\ \emph {et~al.}(2019)\citenamefont {Pavlov},
  \citenamefont {Gurbatov}, \citenamefont {Kudryashov}, \citenamefont
  {Danilov}, \citenamefont {Porfirev}, \citenamefont {Khonina}, \citenamefont
  {Vitrik}, \citenamefont {Kulinich}, \citenamefont {Lapine},\ and\
  \citenamefont {Kuchmizhak}}]{pavlov201910}%
  \BibitemOpen
  \bibfield  {author} {\bibinfo {author} {\bibfnamefont {D.}~\bibnamefont
  {Pavlov}}, \bibinfo {author} {\bibfnamefont {S.}~\bibnamefont {Gurbatov}},
  \bibinfo {author} {\bibfnamefont {S.}~\bibnamefont {Kudryashov}}, \bibinfo
  {author} {\bibfnamefont {P.}~\bibnamefont {Danilov}}, \bibinfo {author}
  {\bibfnamefont {A.}~\bibnamefont {Porfirev}}, \bibinfo {author}
  {\bibfnamefont {S.}~\bibnamefont {Khonina}}, \bibinfo {author} {\bibfnamefont
  {O.}~\bibnamefont {Vitrik}}, \bibinfo {author} {\bibfnamefont
  {S.}~\bibnamefont {Kulinich}}, \bibinfo {author} {\bibfnamefont
  {M.}~\bibnamefont {Lapine}}, \ and\ \bibinfo {author} {\bibfnamefont
  {A.}~\bibnamefont {Kuchmizhak}},\ }\href@noop {} {\bibfield  {journal}
  {\bibinfo  {journal} {Optics letters}\ }\textbf {\bibinfo {volume} {44}},\
  \bibinfo {pages} {283} (\bibinfo {year} {2019})}\BibitemShut {NoStop}%
\bibitem [{\citenamefont {Ivanov}\ \emph {et~al.}(2008)\citenamefont {Ivanov},
  \citenamefont {Rethfeld}, \citenamefont {O'Connor}, \citenamefont {Glynn},
  \citenamefont {Volkov},\ and\ \citenamefont
  {Zhigilei}}]{ivanov2008mechanism}%
  \BibitemOpen
  \bibfield  {author} {\bibinfo {author} {\bibfnamefont {D.~S.}\ \bibnamefont
  {Ivanov}}, \bibinfo {author} {\bibfnamefont {B.}~\bibnamefont {Rethfeld}},
  \bibinfo {author} {\bibfnamefont {G.~M.}\ \bibnamefont {O'Connor}}, \bibinfo
  {author} {\bibfnamefont {T.~J.}\ \bibnamefont {Glynn}}, \bibinfo {author}
  {\bibfnamefont {A.~N.}\ \bibnamefont {Volkov}}, \ and\ \bibinfo {author}
  {\bibfnamefont {L.~V.}\ \bibnamefont {Zhigilei}},\ }\href@noop {} {\bibfield
  {journal} {\bibinfo  {journal} {Applied Physics A}\ }\textbf {\bibinfo
  {volume} {92}},\ \bibinfo {pages} {791} (\bibinfo {year} {2008})}\BibitemShut
  {NoStop}%
\bibitem [{\citenamefont {Vrij}(1966)}]{Vrij1966}%
  \BibitemOpen
  \bibfield  {author} {\bibinfo {author} {\bibfnamefont {A.}~\bibnamefont
  {Vrij}},\ }\href {\doibase 10.1039/DF9664200023} {\bibfield  {journal}
  {\bibinfo  {journal} {Discuss. Faraday Soc.}\ }\textbf {\bibinfo {volume}
  {42}},\ \bibinfo {pages} {23} (\bibinfo {year} {1966})}\BibitemShut {NoStop}%
\bibitem [{\citenamefont {Oron}\ \emph {et~al.}(1997)\citenamefont {Oron},
  \citenamefont {Davis},\ and\ \citenamefont {Bankoff}}]{Oron}%
  \BibitemOpen
  \bibfield  {author} {\bibinfo {author} {\bibfnamefont {A.}~\bibnamefont
  {Oron}}, \bibinfo {author} {\bibfnamefont {S.~H.}\ \bibnamefont {Davis}}, \
  and\ \bibinfo {author} {\bibfnamefont {S.~G.}\ \bibnamefont {Bankoff}},\
  }\href@noop {} {\bibfield  {journal} {\bibinfo  {journal} {Rev. Mod. Phys.}\
  }\textbf {\bibinfo {volume} {69}},\ \bibinfo {pages} {931} (\bibinfo {year}
  {1997})}\BibitemShut {NoStop}%
\bibitem [{\citenamefont {Chen}\ \emph {et~al.}(1991)\citenamefont {Chen},
  \citenamefont {Levi},\ and\ \citenamefont {Tosatti}}]{Hamaker}%
  \BibitemOpen
  \bibfield  {author} {\bibinfo {author} {\bibfnamefont {X.}~\bibnamefont
  {Chen}}, \bibinfo {author} {\bibfnamefont {A.}~\bibnamefont {Levi}}, \ and\
  \bibinfo {author} {\bibfnamefont {E.}~\bibnamefont {Tosatti}},\ }\href
  {\doibase https://doi.org/10.1016/0039-6028(91)91070-E} {\bibfield  {journal}
  {\bibinfo  {journal} {Surface Science}\ }\textbf {\bibinfo {volume}
  {251-252}},\ \bibinfo {pages} {641 } (\bibinfo {year} {1991})}\BibitemShut
  {NoStop}%
\bibitem [{\citenamefont {Seemann}\ \emph {et~al.}(2001)\citenamefont
  {Seemann}, \citenamefont {Herminghaus},\ and\ \citenamefont
  {Jacobs}}]{seemann2001gaining}%
  \BibitemOpen
  \bibfield  {author} {\bibinfo {author} {\bibfnamefont {R.}~\bibnamefont
  {Seemann}}, \bibinfo {author} {\bibfnamefont {S.}~\bibnamefont
  {Herminghaus}}, \ and\ \bibinfo {author} {\bibfnamefont {K.}~\bibnamefont
  {Jacobs}},\ }\href@noop {} {\bibfield  {journal} {\bibinfo  {journal}
  {Journal of Physics: Condensed Matter}\ }\textbf {\bibinfo {volume} {13}},\
  \bibinfo {pages} {4925} (\bibinfo {year} {2001})}\BibitemShut {NoStop}%
\bibitem [{\citenamefont {Krechetnikov}\ and\ \citenamefont
  {Homsy}(2009)}]{Krechetnikov2009}%
  \BibitemOpen
  \bibfield  {author} {\bibinfo {author} {\bibfnamefont {R.}~\bibnamefont
  {Krechetnikov}}\ and\ \bibinfo {author} {\bibfnamefont {G.~M.}\ \bibnamefont
  {Homsy}},\ }\href {\doibase https://doi.org/10.1016/j.jcis.2008.11.079}
  {\bibfield  {journal} {\bibinfo  {journal} {Journal of Colloid and Interface
  Science}\ }\textbf {\bibinfo {volume} {331}},\ \bibinfo {pages} {555 }
  (\bibinfo {year} {2009})}\BibitemShut {NoStop}%
\bibitem [{\citenamefont {Khare}\ \emph {et~al.}(2007)\citenamefont {Khare},
  \citenamefont {Brinkmann}, \citenamefont {Law}, \citenamefont {Gurevich},
  \citenamefont {Herminghaus},\ and\ \citenamefont
  {Seemann}}]{khare2007dewetting}%
  \BibitemOpen
  \bibfield  {author} {\bibinfo {author} {\bibfnamefont {K.}~\bibnamefont
  {Khare}}, \bibinfo {author} {\bibfnamefont {M.}~\bibnamefont {Brinkmann}},
  \bibinfo {author} {\bibfnamefont {B.~M.}\ \bibnamefont {Law}}, \bibinfo
  {author} {\bibfnamefont {E.~L.}\ \bibnamefont {Gurevich}}, \bibinfo {author}
  {\bibfnamefont {S.}~\bibnamefont {Herminghaus}}, \ and\ \bibinfo {author}
  {\bibfnamefont {R.}~\bibnamefont {Seemann}},\ }\href@noop {} {\bibfield
  {journal} {\bibinfo  {journal} {Langmuir}\ }\textbf {\bibinfo {volume}
  {23}},\ \bibinfo {pages} {12138} (\bibinfo {year} {2007})}\BibitemShut
  {NoStop}%
\bibitem [{\citenamefont {Seemann}\ \emph {et~al.}(2011)\citenamefont
  {Seemann}, \citenamefont {Brinkmann}, \citenamefont {Herminghaus},
  \citenamefont {Khare}, \citenamefont {Law}, \citenamefont {McBride},
  \citenamefont {Kostourou}, \citenamefont {Gurevich}, \citenamefont {Bommer},
  \citenamefont {Herrmann} \emph {et~al.}}]{seemann2011wetting}%
  \BibitemOpen
  \bibfield  {author} {\bibinfo {author} {\bibfnamefont {R.}~\bibnamefont
  {Seemann}}, \bibinfo {author} {\bibfnamefont {M.}~\bibnamefont {Brinkmann}},
  \bibinfo {author} {\bibfnamefont {S.}~\bibnamefont {Herminghaus}}, \bibinfo
  {author} {\bibfnamefont {K.}~\bibnamefont {Khare}}, \bibinfo {author}
  {\bibfnamefont {B.~M.}\ \bibnamefont {Law}}, \bibinfo {author} {\bibfnamefont
  {S.}~\bibnamefont {McBride}}, \bibinfo {author} {\bibfnamefont
  {K.}~\bibnamefont {Kostourou}}, \bibinfo {author} {\bibfnamefont
  {E.}~\bibnamefont {Gurevich}}, \bibinfo {author} {\bibfnamefont
  {S.}~\bibnamefont {Bommer}}, \bibinfo {author} {\bibfnamefont
  {C.}~\bibnamefont {Herrmann}},  \emph {et~al.},\ }\href@noop {} {\bibfield
  {journal} {\bibinfo  {journal} {Journal of Physics: Condensed Matter}\
  }\textbf {\bibinfo {volume} {23}},\ \bibinfo {pages} {184108} (\bibinfo
  {year} {2011})}\BibitemShut {NoStop}%
\bibitem [{\citenamefont {Ricci}\ and\ \citenamefont
  {Novakovic}(2001)}]{ricci2001wetting}%
  \BibitemOpen
  \bibfield  {author} {\bibinfo {author} {\bibfnamefont {E.}~\bibnamefont
  {Ricci}}\ and\ \bibinfo {author} {\bibfnamefont {R.}~\bibnamefont
  {Novakovic}},\ }\href@noop {} {\bibfield  {journal} {\bibinfo  {journal}
  {Gold Bulletin}\ }\textbf {\bibinfo {volume} {34}},\ \bibinfo {pages} {41}
  (\bibinfo {year} {2001})}\BibitemShut {NoStop}%
\bibitem [{\citenamefont {Danilov}\ \emph {et~al.}(2016)\citenamefont
  {Danilov}, \citenamefont {Gonchukov}, \citenamefont {Ionin}, \citenamefont
  {Khmelnitskii}, \citenamefont {Kudryashov}, \citenamefont {Nguyen},
  \citenamefont {Rudenko}, \citenamefont {Saraeva},\ and\ \citenamefont
  {Zayarny}}]{danilov2016background}%
  \BibitemOpen
  \bibfield  {author} {\bibinfo {author} {\bibfnamefont {P.~N.}\ \bibnamefont
  {Danilov}}, \bibinfo {author} {\bibfnamefont {S.~A.}\ \bibnamefont
  {Gonchukov}}, \bibinfo {author} {\bibfnamefont {A.~A.}\ \bibnamefont
  {Ionin}}, \bibinfo {author} {\bibfnamefont {R.~A.}\ \bibnamefont
  {Khmelnitskii}}, \bibinfo {author} {\bibfnamefont {S.~I.}\ \bibnamefont
  {Kudryashov}}, \bibinfo {author} {\bibfnamefont {T.~T.}\ \bibnamefont
  {Nguyen}}, \bibinfo {author} {\bibfnamefont {A.~A.}\ \bibnamefont {Rudenko}},
  \bibinfo {author} {\bibfnamefont {I.~N.}\ \bibnamefont {Saraeva}}, \ and\
  \bibinfo {author} {\bibfnamefont {D.~A.}\ \bibnamefont {Zayarny}},\
  }\href@noop {} {\bibfield  {journal} {\bibinfo  {journal} {Laser Physics
  Letters}\ }\textbf {\bibinfo {volume} {13}},\ \bibinfo {pages} {055602}
  (\bibinfo {year} {2016})}\BibitemShut {NoStop}%
\bibitem [{\citenamefont {Paeng}\ \emph {et~al.}(2015)\citenamefont {Paeng},
  \citenamefont {Yoo}, \citenamefont {Yeo}, \citenamefont {Lee}, \citenamefont
  {Kim}, \citenamefont {Ko},\ and\ \citenamefont
  {Grigoropoulos}}]{paeng2015low}%
  \BibitemOpen
  \bibfield  {author} {\bibinfo {author} {\bibfnamefont {D.}~\bibnamefont
  {Paeng}}, \bibinfo {author} {\bibfnamefont {J.-H.}\ \bibnamefont {Yoo}},
  \bibinfo {author} {\bibfnamefont {J.}~\bibnamefont {Yeo}}, \bibinfo {author}
  {\bibfnamefont {D.}~\bibnamefont {Lee}}, \bibinfo {author} {\bibfnamefont
  {E.}~\bibnamefont {Kim}}, \bibinfo {author} {\bibfnamefont {S.~H.}\
  \bibnamefont {Ko}}, \ and\ \bibinfo {author} {\bibfnamefont {C.~P.}\
  \bibnamefont {Grigoropoulos}},\ }\href@noop {} {\bibfield  {journal}
  {\bibinfo  {journal} {Advanced Materials}\ }\textbf {\bibinfo {volume}
  {27}},\ \bibinfo {pages} {2762} (\bibinfo {year} {2015})}\BibitemShut
  {NoStop}%
\bibitem [{\citenamefont {Busleev}\ \emph {et~al.}(2019)\citenamefont
  {Busleev}, \citenamefont {Ivanova}, \citenamefont {Kudryashov}, \citenamefont
  {Rudenko}, \citenamefont {Zayarny},\ and\ \citenamefont
  {Ionin}}]{busleev2019fano}%
  \BibitemOpen
  \bibfield  {author} {\bibinfo {author} {\bibfnamefont {N.}~\bibnamefont
  {Busleev}}, \bibinfo {author} {\bibfnamefont {A.}~\bibnamefont {Ivanova}},
  \bibinfo {author} {\bibfnamefont {S.}~\bibnamefont {Kudryashov}}, \bibinfo
  {author} {\bibfnamefont {A.}~\bibnamefont {Rudenko}}, \bibinfo {author}
  {\bibfnamefont {D.}~\bibnamefont {Zayarny}}, \ and\ \bibinfo {author}
  {\bibfnamefont {A.}~\bibnamefont {Ionin}},\ }\href@noop {} {\bibfield
  {journal} {\bibinfo  {journal} {Plasmonics}\ ,\ \bibinfo {pages} {1}}
  (\bibinfo {year} {2019})}\BibitemShut {NoStop}%
\end{thebibliography}
\end{document}